# Nitrogen and Oxygen Transport and Reactions During Plasma Nitridation of Zirconium Thin Films

L. Pichon [a], A. Straboni [a], T. Girardeau [a], M. Drouet [a], P. Widmayer [b]

[a] *Laboratoire de Métallurgie Physique - Université de Poitiers - UMR 6630 CNRS*
*SP2MI Bld 3 Téléport 2 BP179 86960 FUTUROSCOPE CEDEX   FRANCE*
*phone: (33) 5 49 49 6739   fax: (33) 5 49 49 66 92*
*e-mail: Alain.Straboni@lmp.univ-poitiers.fr*

[b] *Abteilung Festkoerperphysik - Ulm University - GERMANY*

# Abstract

Zirconium nitride ZrN is a refractory material with good mechanical and thermal properties. It is therefore a good candidate for hard surface treatment used at high temperature. In this work, we report the growth and characterisation of ZrN by plasma assisted thermal nitridation of zirconium films in an $NH_3$ atmosphere. The process was monitored by in-situ monochromatic ellipsometry and the grown nitrides were profiled and analysed by Auger Electron Spectroscopy. By using temperatures in the 700 - 800°C range, the obtained material is quite close to ZrN, but, depending on experimental conditions, residual oxygen (impurities) can be easily incorporated by reaction with zirconium. The analysis of the ellipsometric data has shown that the nitridation did not occur by a simple growth of nitride on zirconium. Auger profiles confirmed the presence of an oxidised zirconium layer localised between the nitrided surface and the remaining metal. This oxidation was observed to occur preferentially during the temperature ramping, that is, in the low temperature regime. At high temperature, nitridation is dominant and the incorporated oxygen is exchanged with nitrogen. Oxygen is then partly rejected by diffusion out of the film through the ZrN surface layer and partly by diffusion in the deep zirconium sublayer. By using these observations, a new model of growth with a layered $ZrN/ZrO_x/Zr$ film was used to describe in-situ ellipsometric data. By comparing the pure thermal and the plasma treatments, the advantages of the plasma assisted treatment clearly appear : complete nitridation of the zirconium layer was achieved and the oxygen amounts in the film were substantially reduced.

# 1. Introduction

Due to its high hardness (2500HV) and wear resistance, zirconium nitride (ZrN) has motivated recent experimental research. Indeed, because of a great chemical and thermal stability ($T_m$=2850°C) [1], the mechanical properties can be preserved in high temperature and reactive environments [2]. Moreover, zirconium nitride films offer promising optical applications by using the ZrN stable phase, with optical index very similar to that of the gold (N=0.5-i3.2 at $\lambda$=633 nm), in combination with the transparent $Zr_3N_4$ metastable phase [3, 4]. The Zr:1N:1 stoichiometry is generally obtained when using reactive sputter-deposition by adjusting inert and nitriding gas flows [5]. Native thermal or plasma growth in a reactive atmosphere presents several advantages over deposition such as non directional layer growth, gradual interface between film and substrate, and production of the stable structure. However, the nitride growth resulting from refractory solid/nitriding gas reaction suffers two major draw backs : the nitridation kinetic is very slow and the high solubility of oxygen in the metal makes it very sensitive to residual oxygen [6]. Plasma treatment has been proposed for lowering the nitridation or oxidation temperature of semiconductors, metals or alloys or for enhancing the reaction kinetic [7]. Very few results are reported in the literature on zirconium plasma nitridation [8]. In addition, the effect of oxygen during nitridation is not well understood. The purpose of this work is to explore plasma nitridation of zirconium films on silicon substrates by examining nitrogen and oxygen transport during the process.

# 2. Experimental details

Zirconium 100 nm thin films were deposited by ion beam sputtering [9] on silicon. The substrates were previously nitrided by plasma treatment [10] in order to avoid zirconium-silicon reaction. These films were analysed by Rutherford Backscattering Spectrometry and Nuclear Reaction Analysis and they are almost oxygen free (less than 3 at.%) ; some iron contamination (3 at.%) was detected due probably to sputtering of the chamber walls.

The nitridation treatments of the zirconium film were realised in a plasma assisted thermal reaction chamber previously described [11, 12, 13]. In this reactor, a classical thermal treatment can be performed in combination with plasma treatment in a gaseous atmosphere. The molecules and ions are in thermal equilibrium with the reactor walls ; their kinetic energy is about 0,07 eV whereas the electron temperature is about 3 eV. The nitriding ($NH_3$) plasma was generated at $93.10^{-3}$ mbar with an external capacitively coupled 13.56-Mhz generator. A turbomolecular pump allows a base pressure of about $5.10^{-7}$ mbar. The nitridation process was sequenced as follows: introduction of the sample and pumping to $5.10^{-7}$ mbar - introduction of 80 sccm of $NH_3$ and stabilisation of the pressure at $93.10^{-3}$ mbar - creation of the 700W plasma and stabilisation of the pressure - ramping of temperature to working temperature at time t=0 ; after the desired reaction time, plasma is stopped and $NH_3$ is exchanged with $N_2$ at the same pressure. The samples were pulled outside the hot zone to a cold zone without breaking the vacuum. We studied three treatment temperatures : 500, 700 and 800°C. To compare plasma nitridation with a pure thermal nitridation, treatments were realised at the same low pressure without plasma in the cases of 500 and 700°C.

In order to analyse the nitridation process, two treatments at 700°C were interrupted after 15 minutes and 2 hours. The temperature evolution during temperature ramping is roughly linear with time between Room Temperature (RT) and 700°C, during the first 30 minutes. This means that, after 15 minutes, the sample temperature did not reach 700°C. As a matter of fact, it only reached about 250 - 300°C.

During the treatment, the evolution of the nitrided film optical characteristics was observed using an in-situ monochromatic rotating analyzer ellipsometer ($\lambda$=633nm) [14].

Optical characterisation of nitrided samples were realised ex-situ on a commercially-available variable angle spectroscopic ellipsometer (SOPRA). The measurements were made from 0.2 µm to 1.2 µm at a 75° angle of incidence , which is the most sensitive angle for measurements on silicon substrates, since it is close to the Brewster angle of silicon. For opaque materials (ZrN for example), the film is treated as a semi-infinite bulk material and optical constants can be directly extracted from experimental measurements.

For X-Ray diffraction (XRD) studies, we used an original four circle goniometer [15]. The choice of the incident angle is independent of the 2Θ angle. It allows two kinds of geometry : classic Θ–2Θ and grazing incidence. The linear detector, with a window angle of 17° and a precision of 0.045° could run with a spectral range of between 0 -150° in 2Θ. The spot size on the sample is about $1\times1mm^2$.

Each geometry gives different and complementary information on the film structure. In classic Θ–2Θ symmetry, only the reticular planes parallel to the surface are diffracting. So only textured grains are investigated. In grazing incidence (2-5°), the non-oriented grains are investigated, i.e. the reticular planes which are not parallel to the surface. As there is no preferential orientation in this geometry, a powder like spectrum is usually obtained, with broadened peaks due to a smaller grain size.

Auger Electron Spectroscopy (AES) experiments were performed on an "Omicron" Auger spectrometer system, with a cylindrical mirror analyser and an integrated electron source [16]. The AES spectra were recorded with a 3 keV electron incident beam and the Auger electrons were analysed in the 50-550 eV range. Depth profiles were obtained using different sputtering conditions with a 3 keV or 5 keV argon ion beam. These two conditions of profiling and the nature of the compound (metal, nitride, oxide) lead to a large difference in the time necessary to sputter the whole nitrided film and to reach the interface : about 70 - 80 minutes with 3 keV beam and 20 - 25 minutes with 5 keV beam.

A calibration of nitrogen and oxygen stoichiometry was performed using pure $ZrO_2$ (obtained by thermal oxidation of zirconium layer) and pure ZrN (obtained by reactive ion deposition) layers. The Auger peaks monitored during sputtering are as follows: 120 eV (Zr), 385 eV (N), 510 eV (O), and 90 eV (Si or Zr). In depth N and O concentrations were determined by drawing the N/Zr and O/Zr atomic ratios. Because of the overlapping silicon and zirconium peaks at 90 eV, the ratio of Si/Zr seems to be non-zero in the film. Therefore, its value is meaningless and we only used the increase of the Si/Zr ratio to distinguish the substrate and film interface.

High oxygen concentrations are clearly visible on the top of each sample ; it is due to the very thin (a few nanometers) natural oxide on the nitride surface. Moreover, because of the preferential sputtering of volatile elements by argon ion bombardment, the analysis ratio at the surface region is modified at the beginning of the profile until a stationary state is obtained. Therefore, we have calibrated the O/Zr value only in the stationary part of the profile, that occurs after a few minutes of sputtering.

### 3. Results

#### *3.1 Comparison of plasma and thermal treatments*

The AES profiles of the zirconium films nitrided during 7 hours with or without plasma at 700 and 500°C are presented in figure 1 and figure 2. Concerning the treatment at 700°C with plasma (figure 1b), the composition is almost constant through the film thickness: the zirconium is completely nitrided and the N/Zr level is close to 1. Oxygen is present with a constant level O/Zr of 0,5. On the contrary, the thermally nitrided film (figure 2b) exhibits two layers of different composition. Near the surface, a first layer is composed of a nitride with $ZrNO_{0.3}$ stoichiometry ; this composition is still constant on half of the film. Beneath this layer, the zirconium is fully oxidised with a composition of $ZrO_2N_{0.2}$.

In the case of 500°C treatment, the nitridation is not complete with (figure 1a) nor without (figure 2a) plasma. However, with plasma, a very thin ZrN layer could be observed at the surface. This was not the case with pure thermal treatment where there is only a short gradient of nitrogen from N/Zr equal to 0.8 to 0. The big difference between these two profiles consist in the oxidation level of the sublayer: with thermal treatment, a great part of the film is fully oxidized to $ZrO_2$ stoichiometry and the nitrided surface layer still has a significant oxygen level. On the contrary, the oxidation level is clearly smaller with plasma treatment: the maximum O/Zr value is about 0.9 and only a small part of the film is oxidised.

The role of oxygen seems to be of great importance in the zirconium nitridation process ; the great quantity fixed in the film is due to both the oxidising species present in the residual atmosphere (leaks, out-gassing, and reduction of the silica tube by the $NH_3$ plasma) and the very high affinity of zirconium with oxygen, even at low temperature.

By comparing the AES profiles at higher temperature, plasma treatment appears to produce an improvement in the nitridation efficiency. Firstly, the nitridation kinetic is faster ; 7 hours at 700°C (figure 1b) is sufficient to completely nitridate a 100 nm thick zirconium film, compared to thermal treatment where only half the film is nitrided (figure 2b). Secondly, the thermal process leads to the formation of a highly oxidised zirconium sublayer, with a stoichiometry closed to $ZrO_2$, whereas the plasma-assisted process substantially reduces oxygen incorporation.

*3.2 Influence of the temperature*

Considering the AES profiles of the films nitrided in plasma at 700 and 800°C as shown in figure 1, it can be seen that the nitrogen and oxygen compositions are constant. The zirconium is completely nitrided and the N/Zr level is close to 1 throughout the thickness of the film. Oxygen is minimised at 800°C with a constant value of the O/Zr ratio of about 0.2.

The plasma nitridation efficiency is observed to be higher when the temperature is increased. Naturally, the nitrogen diffusion is accelerated and, with a temperature as low as 700°C, the film can be completely nitrided in less than 7 hours of treatment. Moreover, high temperature conditions lead to a lowering of the final oxidation level. This is probably due to the more efficient reaction of nitrogen with zirconium at these temperatures.

### *3.3 Characterisation of the 700 and 800°C plasma nitrided films*

XRD experiments were performed in order to study the microstructure of the ZrN layers and the oxygen states in the film (interstitial, oxide, etc...). The spectra recorded in the [20° - 45°] 2Θ range in symmetric and grazing incidences are presented in figure 3 for the two samples. A NaCl structure characteristics of ZrN appears through the peaks (111) and (200) under the grazing incidence conditions. The obtained lattice parameter calculated from these spectra is quite closed to that of bulk ZrN one (0.458 nm). In symmetric geometry, such a ZrN phase does not appear. On the contrary, the observable peaks can be related to zirconium oxide phases ($ZrO_2$ and $Zr_3O_{1-x}$), and this indicates that some oxygen was fixed in the layer as crystallised oxide grains.

Optical characterisations of the two samples were performed by spectroscopic ellipsometry in the 0.2-1.2 µm spectral range. The evolution of the refractive index and of the extinction coefficient with wavelength are reported in figure 4; they are compared with those of a sputter-deposited ZrN layer obtained by dual ion beam sputtering, showing a low level of oxygen contamination (<2%) [17]. All the experimental curves could be fitted using an optical model based on the free-electron Drude model and a bound-electron Lorentz oscillator with an absorption wavelength of about 0.22 µm. The parameters deduced from the model, such as plasma energy and relaxation energy of conduction electrons, are presented in table I together with data extracted from the literature [18, 19]. The measured density of free electrons was observed to be higher in the 800°C treated sample than in the 700°C treated sample, however, it always remains slightly lower than for pure ZrN. This means that some of the zirconium atoms are bound to oxygen resulting in trapping of electrons in Zr-O bonds. Moreover, the relaxation energy shows that the life-time of the conduction electrons was significantly shorter in the nitrided layers than in the sputter-deposited ZrN films. As the grain size was comparable in both materials, the observed difference may be related to the presence of additional oxygen in interstitial sites in the ZrN lattice or to the presence of oxide crystallites. The model, which is well suited for the ZrN reference, does not fit very well the refractive index behaviour of our two samples with wavelength. It is another indication of the occurrence of some additional optical effects due to, for example, the presence of oxides which were not taken into account by the model.

## 4. Study of the nitridation mechanisms

### *4.1 AES profiling of intermediary samples*

In order to understand the nitridation process, a chemical analysis by AES Profiling was performed on films nitrided at 700°C for both 15 minutes and 2 hours. The spectra obtained are presented on figure 5. The profile of the sample treated for 15 minutes (figure 5a) shows that nitrogen has just started to diffuse from the zirconium surface. A short nitrogen gradient follows. One can notice that, behind the nitride layer, an oxide layer is present with a maximal composition $ZrO_{1.3}$ and with a diffuse interface due to a deeper oxygen diffusion gradient. More deeply in the layer, zirconium is unreacted. For longer treatment time, as shown in the profile of the plasma-nitrided sample at 700°C during 2h (figure 5b), the whole film has been nitrided with a small in-depth gradient of nitrogen concentration ranging from ZrN to $ZrN_{0.5}$ composition. Near the surface, the oxygen concentration is constant with a composition about O/Zr= 0.25 whereas, under the nitride layer and till the substrate, it increases to a maximum O/Zr=0.7. After 7 hours of treatment (figure 1b), the completely nitrided zirconium layer is homogeneous on the whole depth with a final composition of $Zr_1N_1O_{0.5}$.

The evolution of oxygen-nitrogen profiles with time show that the zirconium reaction proceeds through a competition between nitridation and oxidation. It seems that oxidation occurs mainly at the beginning of the treatment; oxygen is thought to be incorporated during temperature ramping. So, at low temperature, oxidation of zirconium by residual oxygen predominates through a rapid diffusion in the film. When the temperature is high enough, nitrogen starts to react at the surface with the partially oxidised zirconium. Together with nitrogen diffusion, exchange reactions take place; nitrogen is incorporated in the oxidised region whereas oxygen is rejected.

*4.2 In-situ ellipsometric study of the nitridation process*

The evolution of the ellipsometric parameters (tan(Ψ), cos(Δ)) measured on the sample plasma nitrided at 700°C for 7 hours is presented in figure 6 (triangles, squares, circles). The first point (t=0) corresponds to the non reacted zirconium layer, on a thin silicon nitride (2 nm), on a silicon substrate. The final point (t=7 hours) corresponds to the final obtained material $ZrNO_{0.5}$, with N=1.63-i1.89. On the basis of the Auger profile analysis, a layered model was used for fitting the ellipsometric data with a layered $ZrN/ZrO_x/Zr/Si$ structure, with the different layer thicknesses increasing with time. The first step of the growth process is believed to begin with oxidation which consumes zirconium and is followed by nitridation, consuming the oxide, until the metal/silicon interface is reached. To take into account the oxygen and nitrogen gradients, the experimental curve has been fitted with a structure consisting of 9 layers (figure 6 in grey). Their optical properties are summarised in the table II.

In order to compare the simulation results with those obtained by AES profiling, relationships between the value of the optical index and nitrogen or oxygen compositions are needed. The metallic zirconium index was found to be about 2.2 -i3.2, that of ZrN is about 0.5 - i3.2, and that of $ZrO_2$ about 2.2 -i0. Zr and ZrN are metallic with high extinction coefficients, whereas the oxide is transparent. A decrease in the extinction coefficient of the growing layer was interpreted as an increase in the oxygen concentration. In the same way, as the refractive indices of Zr and $ZrO_2$ are much higher (2.2) than that of ZrN (0.5), a decrease in the refractive index was interpreted as an increase in the nitrogen concentration. Assuming these two behaviours, it was possible to simulate the evolution of the oxygen and nitrogen profiles in the film, as shown in figure 7. This figure also shows the evolution of the different thickness versus treatment times chosen in the simulation.

The growth reaction can be divided into three steps. During the first step (figure 7(a)), between 0 and about 11 minutes (i.e. between RT and 200°C), the ellipsometric trajectory ($\tan(\Psi)$ versus $\cos(\Delta)$) follows a curve characteristic of an oxygen gradient progression in the zirconium layer. This means that the layer growing at the surface has its composition closer and closer to $ZrO_2$ (the refractive index increases to about 2.2 and the extinction coefficient vanishes). The most transparent layer has been shown to present an optical index of about 2.25 - i0.85.

During the second step (figure 7(b)), between 11 and 23 minutes (i.e. between 200 and 600°C), the surface layer gradually shifts from dielectric to metallic, with a lower refractive index and an higher extinction coefficient. This step can be interpreted as the formation of a zirconium oxynitride with the nitrogen concentration increasing with time. The oxygen is then replaced by nitrogen. It can be either released out of this surface layer or it can diffuse more deeply into the film and oxidise the remaining zirconium. The nitrogen-rich film at the surface has the optical index characteristics of a metal, about 1.55-i2.4. However, this value is quite different to that of ZrN (0.5-i3.2), which means that a significant amount of oxygen is still present in the nitrided layer. Moreover, to obtain a good fit, an uninterrupted progression of the « oxide » layers below the nitride ones had to be taken into account in the model. As a matter of fact, AES profiles on intermediate samples (15 minutes and 2 hours) confirm that the nitride layer is unable to stop the oxygen diffusion.

The third step (between 23' and 7 hours, i.e. at temperature T=700°C - figure 7(c)) is simulated by the growth of a layer with a slightly higher refractive index and a lower extinction coefficient (1.86-i2.15) than the previous one (1.55-i2.4). If the previously described relationships between optical indices and compositions are applied, then the oxynitride at the surface is less nitrogen-rich and the oxygen content is higher. A comparison of the AES profiles of samples nitrided at 700°C for 2 and 7 hours seems to confirm this.

It is difficult to account for the lower nitridation efficiency during this high temperature step. This result can be interpreted as an homogenization of the oxygen distribution, and possibly as an evolution of the oxygen state in the layer. At the end of the second step (figure 7c - t=22.8 min), all the zirconium atoms at the surface are nitrided while those deeper in the film are all oxidised. During the following nitridation reaction of the deep zirconium oxide, oxygen atoms are continuously replaced by nitrogen and released into the film. As there is no more metallic zirconium to be oxidized, oxygen must then diffuse through the ZrN film to the surface. The tan(Ψ) evolution with time (figure 8) confirms that, compared to the first two steps, the third one concerns a very slow phenomenon. Moreover, the evolution of the ellipsometric parameter tan(Ψ) is obviously parabolic with time (i.e. linear with $t^{1/2}$). Such a parabolic time dependence is characteristic of a diffusion limited mechanism whereas nitridation of the oxide, which is limited by interface reactions, would have led to a linear time dependence. Either the oxygen diffusion or the nitrogen diffusion through the ZrN surface layer can be this limiting mechanism. This analysis is in good agreement with the observation of a surface compound which is slightly more transparent because of its slowly increasing oxygen content, coming from the bulk film nitridation. Furthermore, the mean oxygen content in the film tends to decrease continuously and to become homogeneous, as indicated by the AES profiles in figures 1 and 5. On the other hand, some oxides are thought to grow during this step by the clustering of micro-precipitates. According to XRD results, better crystallized and bigger zirconium oxide grains have been observed in samples treated for longer periods of time. Such oxygen segregation is interpreted as nitride phase and oxide phase separation.

## **Discussion**

The nitridation of zirconium thin films in an ammonia plasma leads to the formation of zirconium nitrides with final compositions of approximately $ZrNO_{0,2}$. High temperature treatments for long periods of time (several hours) are necessary to produce such a layer on the whole 100 nm thick film. XRD analysis shows the presence of several crystallographic phases: the ZrN NaCl-phase and one or two zirconium oxide phases. The optical properties are typical of a metallic material. The free electron density and life time have been found to be slightly lower than in pure ZrN; this is attributed to the presence of oxides and oxygen in interstitial sites of the ZrN lattice.

In situ ellipsometric measurements and AES profiling have enabled the development of a growth model. The gas/metal reaction has been shown to occur via a competitive process between nitridation and oxidation due to residual impurities such as $O_2$ and $H_2O$. At low temperature, thermal activation is not sufficient to allow nitride formation by nitrogen diffusion in the film. On the contrary, residual oxygen reacts easily with zirconium by diffusing in the metal. At higher temperature, an exchange reaction of oxygen with nitrogen occurs and a zirconium nitride layer grows on top of the film. At the same time, the oxidation of the underlying zirconium goes on with the released oxygen. A third step starts at high temperature, when all the zirconium atoms are either nitrided or oxidized. Most of the oxygen then leaves the film by diffusion, leading to a slight increase in the O/Zr level in the ZrN surface layer, but some of it is fixed as crystallised oxides or in interstitial sites. The elimination of oxygen is very slow during this step and needs as long as 7 hours to produce a nitride with a low oxygen concentration. The kinetic limiting mechanisms can be considered to be the diffusion of oxygen and nitrogen-rich species in opposite directions.

The exchange of oxygen with nitrogen, observed during treatments with and without plasma, should happen by the following reaction:

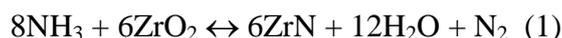

$$8NH_3 + 6ZrO_2 \leftrightarrow 6ZrN + 12H_2O + N_2 \quad (1)$$

However, such reaction is not exothermic and, in the case of thermal treatment, the oxide should then remains unaltered. To understand why ZrN can be produced even without plasma, an oxidation of the remaining metallic zirconium deeply in the film has to be taken into account. This oxidation reaction, as $Zr + xO \rightarrow ZrO_x$ (2), is very exothermic and leads to an important driving force for the transport of oxygen or hydroxy-species from the $ZrN/ZrO_2$ interface towards the $ZrO_2/Zr$ one. The thermal nitridation of the oxide is then made possible as long as some zirconium remains to fix the released oxygen. This mechanism implies that the nitridation must stop when all the zirconium atoms are either oxidised or nitrided. The film should then be composed of two layers of ZrN and $ZrO_2$. Indeed, this bilayer structure is effectively observed on the AES profile of the sample nitrided at 700°C without plasma. Moreover, the in situ ellipsometry measurement on this sample shows that the $(\Psi,\Delta)$ values are stabilised during the last hours of the treatment ; this is another indication consistent with this $ZrN/ZrO_2$ saturation bilayer.

The use of a reactive plasma atmosphere leads to an enhancement of the treatment efficiency in two ways. Firstly, the kinetic of the nitridation process is largely accelerated and the plasma allows a stoichiometric nitride formation in the whole film. Secondly, the oxidation process is reduced and a lower amount of oxygen is finally fixed in the film. During a plasma treatment, the nitridation reaction does not involve only $NH_3$ molecules. The mechanism previously described for thermal nitridation of the oxide (reactions (1) and (2)) is still working but others reactions can occur. Indeed, a high density of excited and unstable species, like ions and radicals ($NH_x^*$, $H_2^*$, $H^+$, $N^*$,...), is produced in the plasma which then constitutes a more powerful reactive atmosphere than the neutral ammonia gas. Moreover, because of their acceleration in the plasma sheath, the ionic species impinging on the surface have some kinetic energy, about 15 eV ($5kT_e$), favouring the activation of surface reactions. Due to the high reactivity of these unstable species, reaction like

$$NH_x^* + ZrO_2 \leftrightarrow ZrN + x/2\ H_2O + (1-x/4)\ O_2 \qquad (3)$$

is thermodynamically more favourable. Besides, the high density at the surface of atomic hydrogen and hydrogen radicals probably plays an important role in the diffusion of oxygen out of the film. Oxygen reacts easily at the surface with hydrogen bearing species to produce water and other hydroxy-species which are then eliminated by the pumping. This consumption of oxygen combined with the high partial pressure of $NH_x$ species (compared to the partial pressures of the oxidising species) is large enough to unbalance the reaction (3) to the right side and to improve the exchange of oxygen with nitrogen in the film.

So, the process previously invoked for the exchange of oxygen with nitrogen is accelerated when using plasma activation. Moreover, unlike thermal treatment, when all zirconium atoms have reacted either with oxygen or nitrogen, the oxygen / nitrogen exchange reaction is going on. The process, which is then limited by the species diffusion, is slow but it leads finally to the complete nitridation of the whole film.

## 4. Conclusion

We have studied the growth of zirconium nitride in an $NH_3$ plasma on Zr layers sputter-deposited on silicon. By in-situ monitoring of the optical parameters of the growing film, we have shown that ellipsometry measurements cannot account for the growth of ZrN on a pure Zr film. This is due to the presence of a nitrogen gradient between the ZrN layer and the Zr metal and to the partial oxidation of the zirconium subsurface by residual oxygen. The occurrence of competitive transport and reactions between nitrogen and oxygen has been shown by Auger profiling. The evolutions of the N and O concentration profiles with time suggest that oxygen is incorporated at low temperature during the heating step. After this initial regime and at temperatures ranging from 700-800°C, plasma assisted nitridation leads to an exchange reaction mechanism where oxygen is replaced by nitrogen and zirconium nitride films grow from the outer surface. A model using a layered ZrN/ZrOx/Zr structure, with the different layer thicknesses varying with time, was shown to fit the general behaviour of the ellipsometry data quite well. The kinetics of plasma nitridation is very slow, once part of the film is converted at the surface to a nearly stoichiometric ZrN, while a $ZrO_x$ film was formed up to the interface with the substrate. The time-limiting mechanism is probably the diffusion of nitrogen through the ZrN surface layer or the diffusion of the released oxygen out of the film. In the pure thermal process, the oxygen / nitrogen exchange is thought to be in direct relation with the oxidation of the remaining zirconium and the final result of the treatment is a $ZrN/ZrO_2$ bilayer. On the contrary, excited species produced in the plasma enhanced the nitridation reaction and the oxygen / nitrogen exchange, which allows a complete nitridation of the film. Hydrogen bearing species produced in the $NH_3$ plasma are thought to play a major role in the enhancement of the oxygen diffusion out of the film. In order to clarify the role of H-species, some experiments using various $N_2$-$H_2$ compositions will be carried out. This work also suggests that it must be possible to produce, by plasma nitridation, the conversion of $ZrO_2$ into ZrN resulting in a conductive layer of ZrN upon an insulating $ZrO_2$ layer.

# Figure captions

Figure 1 :  AES profiles of samples nitrided at 500°C, 700°C and 800°C for 7h with plasma

Figure 2 :  AES profiles of two samples nitrided at 700°C and 500°C for 7h without plasma

Figure 3 :  XRD spectra obtained in symmetric and grazing incidence on the samples nitrided at 700 and 800°C with plasma

Figure 4 :  Comparison of refractive index and extinction coefficient of samples nitrided for 7h with plasma at 800°C (squares) and 700°C (triangles) with reference ZrN ones (circles). Simulation with (Drude + Lorentz oscillator) model (continuous lines).

Figure 5 :  AES profiles of samples nitrided at 700°C with plasma for 15' and 2h

Figure 6 :  Evolution of the monochromatric ellipsometry parameters ($\tan(\Psi)$, $\cos(\Delta)$) measured in situ during plasma treatment of sample nitrided at 700°C, with plasma, for 7h, and its simulation by a 9-layers film (continuous gray line). Step 1 : oxidation T<200°C - 0<t<11.1 min (triangles). Step 2 : nitridation of the oxide 200°C<T<700°C – 11.1 min<t<22.8 min (squares). Step 3 :

composition homogenization and phases separation T=700°C - t>22.8 min (circles)

Figure 7 :   Qualitative evolution of the N/Zr and O/Zr atomic ratio profiles obtained by the ellipsometric simulation - case of sample nitrided at 700°C, with plasma for 7h.

7(a) : Step 1 - T= R.T. to 175°C - t = 0 to 11.1 min

7(b) : Step 2 - T= 175 to 600°C - t = 11.1 min to 23 min

7(c) : Step 3 - T= 700°C - t = 23 min to 7 hours

Figure 8 :   Evolution of Tan ($\Psi$) with square root of time measured by in situ ellipsometry during the plasma treatment at 700°C for 7h.

Table I : Plasma energy $\eta\,\omega_p$, conduction electron density $n_e$, life time of conduction electrons $\tau$ and static conductivity $\sigma_0$ obtained by optical index fit

| Samples | ZrN (ref. [17]) | ZrN (ref. [18]) | ZrN (ref. [19]) | 800°C - 7h - plasma on | 700°C - 7h - plasma on |
|---|---|---|---|---|---|
| $\eta\,\omega_p$ (eV) | 7.27 | 7.3 to 7.8 | 7.17 | 6.99 | 6.73 |
| $n_e$ (.$10^{22}$/cm$^3$) | 3.82 | 3.88 to 4.15 | 3.81 | 3.54 | 3.27 |
| $\eta/\tau$ (eV) | 0.57 | 0.53 to 0.62 | 0.35 | 1.17 | 1.54 |

Table II :   Parameters of the optical structure used to fit the ellipsometric data.

| Layers | Time of growth start (minutes after the treatment beginning) | Temperature at the starting time (°C) | n | k |
|---|---|---|---|---|
| A | 0 | 25 | 2.2 | 2.1 |
| B | 9.5 | 160 | 2.2 | 1.4 |
| C | 10 | 170 | 2.25 | 0.85 |
| D | 11.1 | 175 | 2.25 | 1.1 |
| E | 12 | 190 | 2.2 | 1.3 |
| F | 12.53 | 200 | 2.18 | 1.65 |
| G | 13.3 | 250 | 1.83 | 2.1 |
| H | 14.9 | 300 | 1.55 | 2.4 |
| I | 22.8 | 600 $\Rightarrow$ 700 | 1.86 | 2.15 |

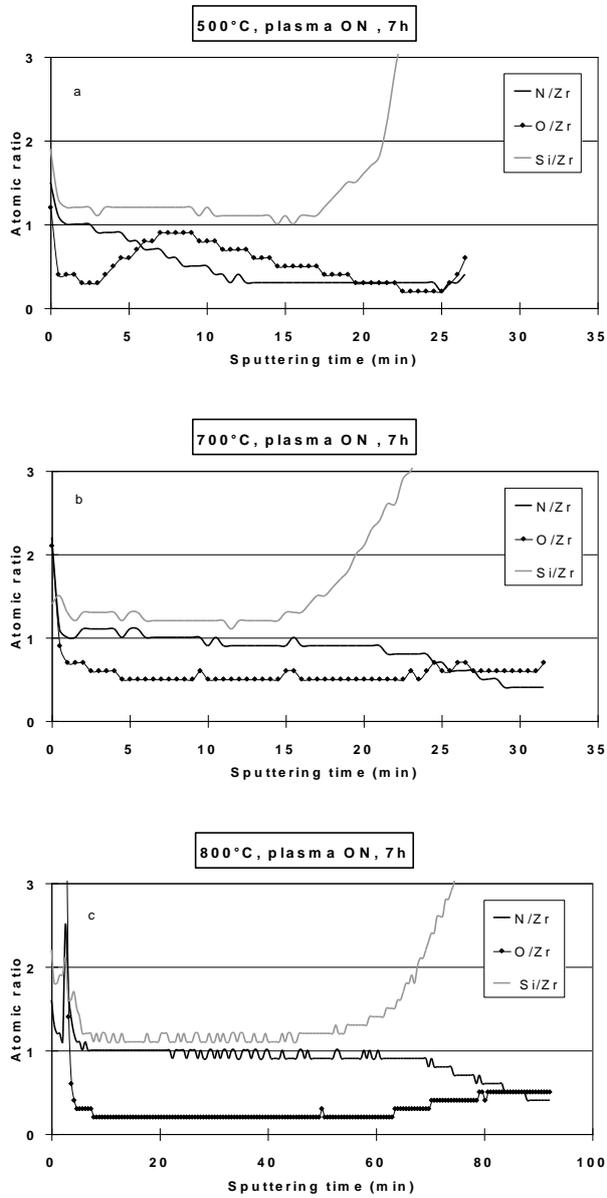

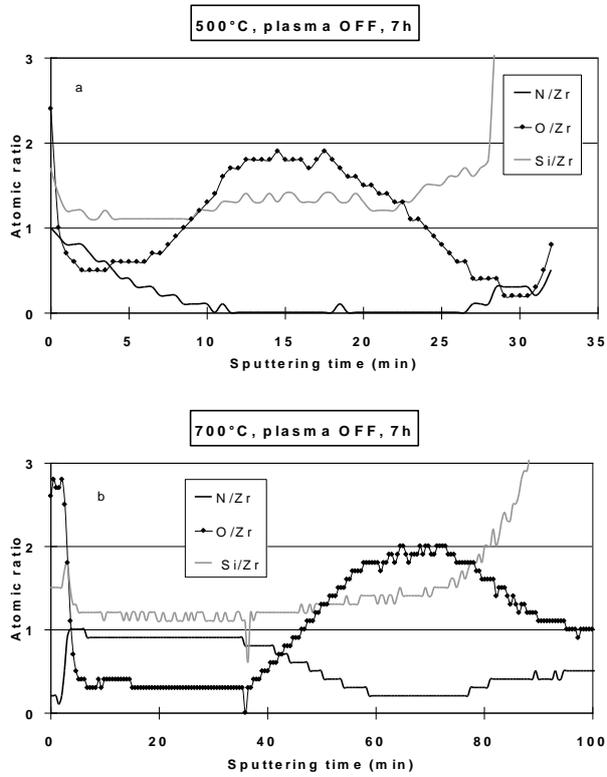

Figure 3 L. PICHON Journal of Applied Physics

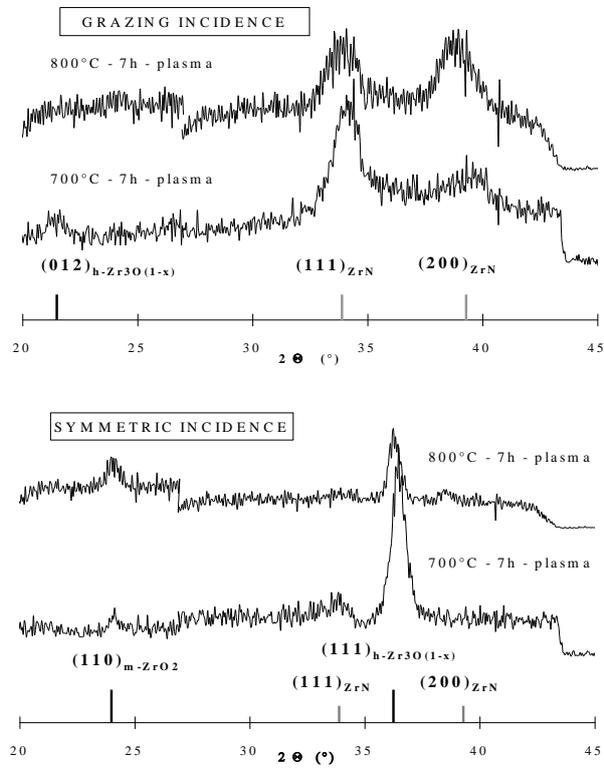

Figure 4    L. PICHON    Journal of Applied Physics

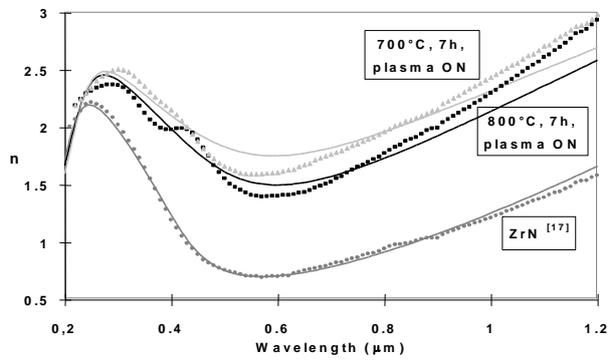

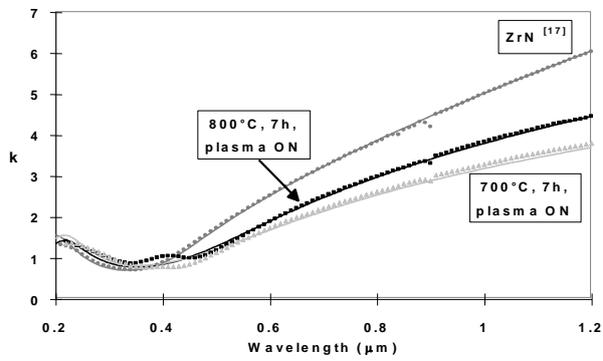

a

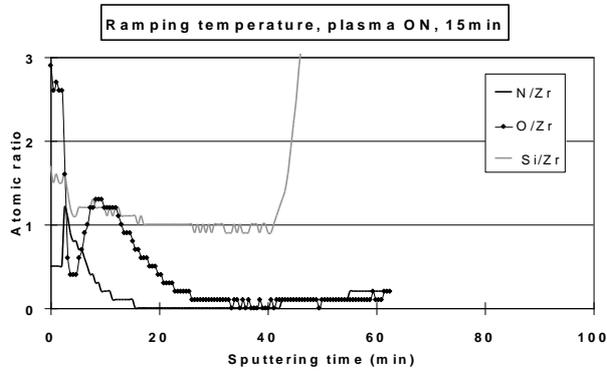

b

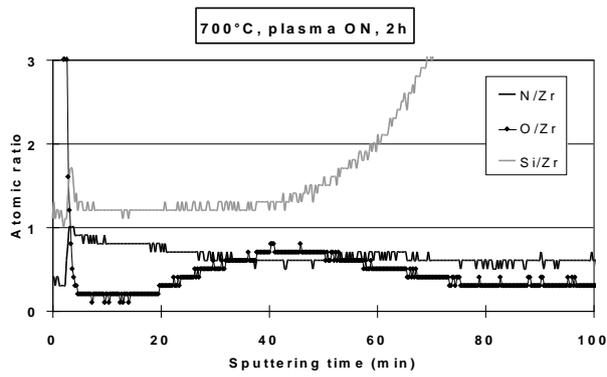

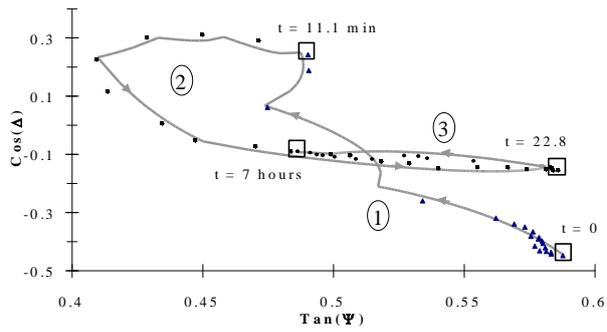

Figure 7 L. PICHON Journal of Applied Physics

8(a)

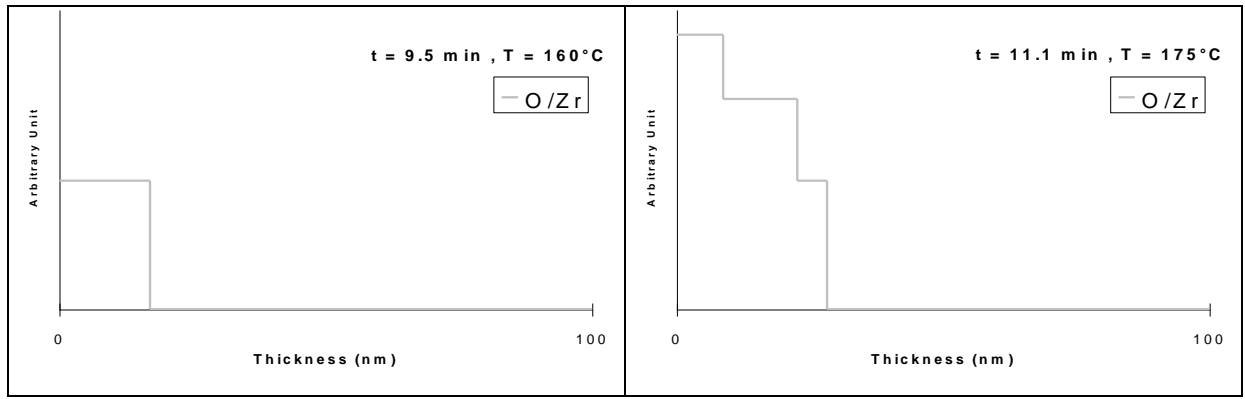

8(b)

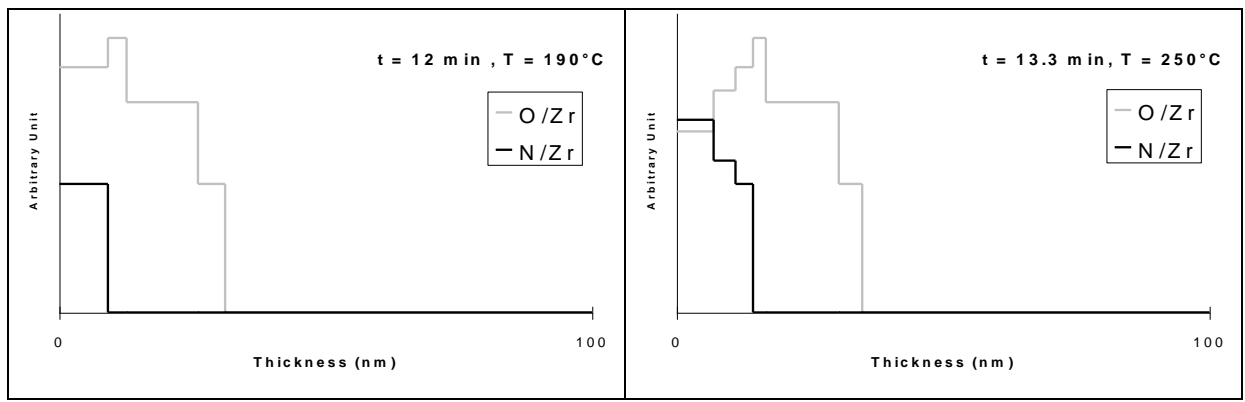

8(c)

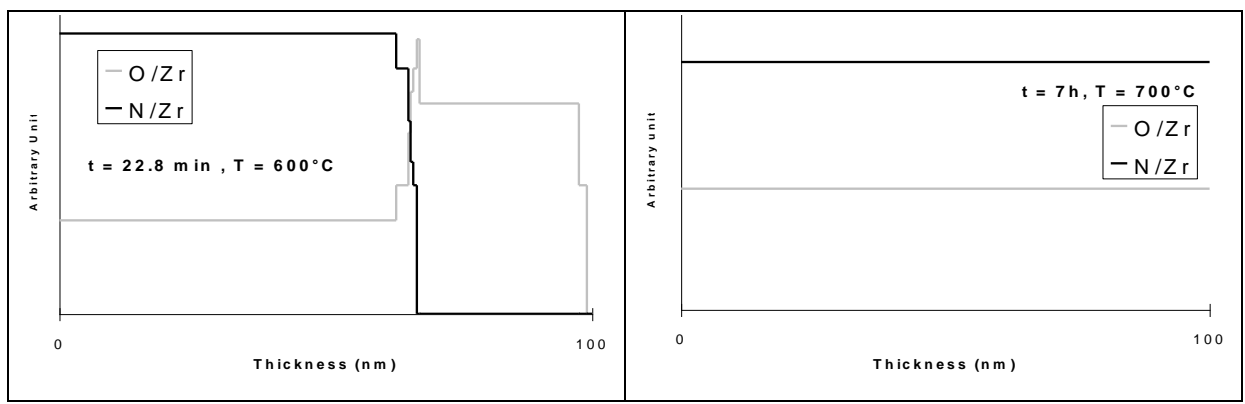

Figure 8  L. PICHON  Journal of Applied Physics

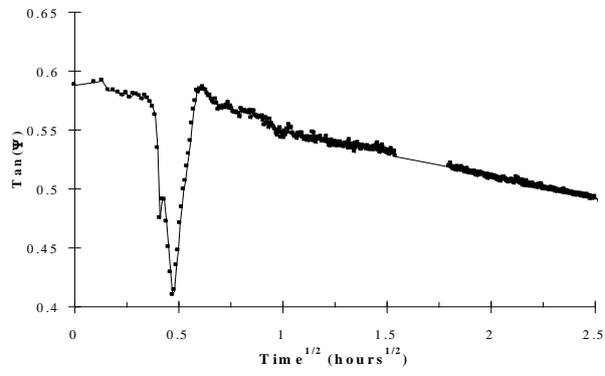